\documentclass[fleqn,10pt]{wlscirep}
\usepackage[utf8]{inputenc}
\usepackage[T1]{fontenc}
\usepackage{url}
\usepackage{graphicx}
\usepackage{subfig}
\usepackage{floatrow}
\usepackage{cases}
\newcommand{\sgn}{\operatorname{sgn}}

\title{Financial Crisis in the Framework of Non-zero Temperature Balance 
Theory}

\author[1,+]{MohammadReza Zahedian}
\author[1,+]{Mahsa Bagherikalhor}
\author[2]{Andrey Trufanov}
\author[1,2,*]{G. Reza Jafari}
\affil[1]{Shahid Beheshti University, Physics Department, Tehran, Iran}
\affil[2]{Institute of Information Technology and Data Science, Irkutsk 
National Research Technical University, 83, Lermontova St., 664074 
Irkutsk, Russia}
\affil[*]{g\_jafari@sbu.ac.ir}
\affil[+]{These authors contributed equally to this work}

\begin{abstract}
Financial crises are known as crashes that result in a sudden loss of value of financial assets in large part and they continue to occur from time to time surprisingly. In order to discover features of the financial network, the pairwise interaction of stocks has been considered in many research, but the existence of the strong correlation of stocks and their collective behavior in crisis made us address higher-order interactions. Hence, in this study, we investigate financial networks by triplet interaction in the framework of balance theory. Due to detecting the contribution of higher-order interactions in understanding the complex behavior of stocks we take the advantage of the orders parameters of the higher-order interactions. Looking at real data of financial market obtained from $S\&P500$ through the lens of balance theory for the quest of network structure in different periods of time near and far from crisis reveals the existence of a structural difference of the network that corresponds to different periods of time. Here, we address two well-known crises the Great regression (2008) and the Covid-19 recession (2020). Results show an ordered structure forms on-crisis in the financial network while stocks behave independently far from a crisis. The formation of the ordered structure of stocks in crisis makes the network resistant against disorder. The resistance of the ordered structure against applying a disorder (temperature) can measure the crisis strength and determine the temperature at which the network transits. There is a critical temperature, $T_{c}$, in the language of statistical mechanics and mean-field approach which above, the ordered structure destroys abruptly and a first-order phase transition occurs. The stronger the crisis, the higher the critical temperature.

\end{abstract}
\begin{document}

\flushbottom
\maketitle
\thispagestyle{empty}


\section*{Introduction}
Financial crises that are a period of reduced economic activity and 
ongoing hardships consist of a sharp drop in international trade and 
rising unemployment and it has happened for as long as the world has had 
currency. Taking advantage of theories to model the financial network to figure out how financial crises develop and how they could be prevented, faces us with a complex system which is not easily predictable. In the last century, numbers of 
market crashes have happened all 
over the world, but there are some important financial crises that have 
impacted on people’s life, namely, the Great Depression in 1929 
\cite{kindleberger_1987} and the Great Recession in 2008 
\cite{united2012world}. Market crashes have been investigated with the 
aim of understanding\cite{shirazi2017,soros} and 
predicting\cite{chatzis}. In \cite{e22091038} authors have addressed the 
collective behavior of markets in the crisis and they benefited the Vicsek 
model to represent an explanation for this complexity. Due to answer to a 
relevant question that how government can play a vital role in economic 
recovery, Bahramin \textit{et. al}\cite{bahrami} used the Ising model to 
analysis the 
hysteresis of the economic network and they suggested a threshold that the 
recovery bill should be bigger than that. In addition, one can construct 
signed weighted network of financial market using correlation matrix 
\cite{mantegna,millington,lim} and interaction matrix \cite{millington} by 
the help of huge economic data stored in the last decades. In the context 
of financial market 
network, the nodes and links represent 
companies and their connections, respectively. The sign and weight of each 
link show the type and strength of  two connected companies. The 
constructed network can be analyzed by different models such as 
percolation\cite{shirazi2017} and random theory 
\cite{jafari2011}. These models can help us to find new features 
and reveal hidden pattern in data. There are many 
investigations in different 
branches of science from financial network \cite{lim,manavi} to 
biological network \cite{schneidman,masoomy} and brain network 
\cite{saberi2021} that consider only pairwise interactions in order to 
find the hidden features of a complex system. Besides all efforts and 
dedicated methods that address crisis, it seems that it 
is still unforeseeable; this research aims to study financial 
network in the framework of balance theory to address higher order interaction in the quest of discovering whether considering higher order interaction can differentiate the structure of the financial network near and far from the crisis? Or how the structure of the network can be affected by crisis?

An outstanding theory that goes beyond pairwise interactions and considers 
higher-order ones is the Heider Balance theory that has been initially 
proposed by Fritz Heider who used it in the psychological science 
\cite{heider}. Then the theory extended to signed networks by Cartwright 
and Harary \cite{cartwright} and applied to other fields such as politics 
\cite{belaza}, ecology\cite{saiz}, and social media \cite{facchetti}. 
Balance theory is based on triadic groups of relationships in which each 
two individuals have an animosity $(-1)$ or a friendship $(+1)$ 
relationship. By considering three-body interactions and the two 
possibility of sign of each relation, four different configurations for 
sign of triangles can occur as follows: 
the triads with three positive relationship ($+++$), two positive 
relationship ($++-$), one positive relationship ($+--$) , and three 
negative relationship ($---$). Triads with 
even (odd) number of negative relationships are called balanced 
(unbalanced). The main purpose of the balance theory is decreasing the 
tension and lessening the number of unbalanced triads in a network of 
relation and moving toward the balance state. Balanced state of a network 
occurs if and only if the all sign of relations in the network be positive 
(heaven) or two cliques create that individuals in each clique have 
positive relations and all of relations between two cliques are negative 
(bipolar). Dynamics of balance theory probes how a network starting from a random initial state moves to the heaven or bipolar\cite{antal1,antal2,estrada2019}.

Authors in \cite{antal1} and\cite{kolakowski} 
proposed the discreet time and continuous time model, respectively, to 
study the dynamics of the balance theory on a fully connected network. The dynamic of a network of three-body interaction in nonzero temperature 
has studied in\cite{bagheri} that uses the balance theory in a fully 
connected network with weighted triangles which weights are coming from Normal distribution.
Marvel \textit{et al.} \cite{marvel} defines energy landscape for networks 
to reveal that networks may be trapped in the local minima so-called, 
jammed states, which their structure depends on the size of the network. Belaza \textit{et al.} investigate the balance theory from statistical 
physics point of view\cite{belaza} and model the triadic energy and tested it 
on data set of relations between countries during the Cold War era. Balance theory also manged to find application in brain network to differentiate the brain networks of participants with autism spectrum disorder from healthy control\cite{zahra}.
Besides, investigations on balance theory that focusing on the 
properties of the real world data; Estrada \textit{et al.}\cite{estrada2014} introduced a method to measure the degree of balance and find that social networks have low degree of balance. Also\cite{facchetti} calculate the global level of balance of very large online social network. 

While there are many investigations on financial markets 
\cite{manavi,millington,namaki2011}, understanding financial crisis using 
market's network under higher order interaction is the main purpose of 
this study. In fact we are looking for hidden structure of a market by 
going beyond pairwise interactions in order to study the collective 
behavior of markets. To this end, we use models of statistical 
physics that worked in the past properly 
\cite{bagheri,amir1,masoumi}. 
Actually, we believe that 
understanding of the financial markets is beyond the simple averaging of 
indexes such as the overall index and the annual return. In this paper, we 
aim for understanding of the financial markets and financial crisis using 
market network. To do so, we specify crisis periods (on-crisis) and normal 
periods (off-crisis) in the observational data and model the market 
network (MN) by triadic interactions of companies, forming triangle 
interactions, then we utilize the balance theory on the MN to investigate 
the 
behavior of the market in different periods of time, and compare results 
of crisis periods and normal periods. We calculated the balance theory’s 
statistical parameters as a function of temperature and analyzed them from 
the financial point of view. The results show there is a big difference 
between on-crisis and off-crisis periods in some characteristic such as 
the number of balanced triads, energy of the network and the critical 
temperature. We use 40 time series of $S\&P500$ companies. Using these 
time series and their correlations, we construct financial network of the 
market in several time windows. After the construction of the financial 
network, we run the balance theory on the correlation network of stocks and we seek 
the dynamics of the constructed network in non zero temperature.

\section*{From Data Description to Network Construction}
\begin{figure}[h]
	\centering
	\hspace{-0.3cm}  
	\includegraphics[scale=.175]{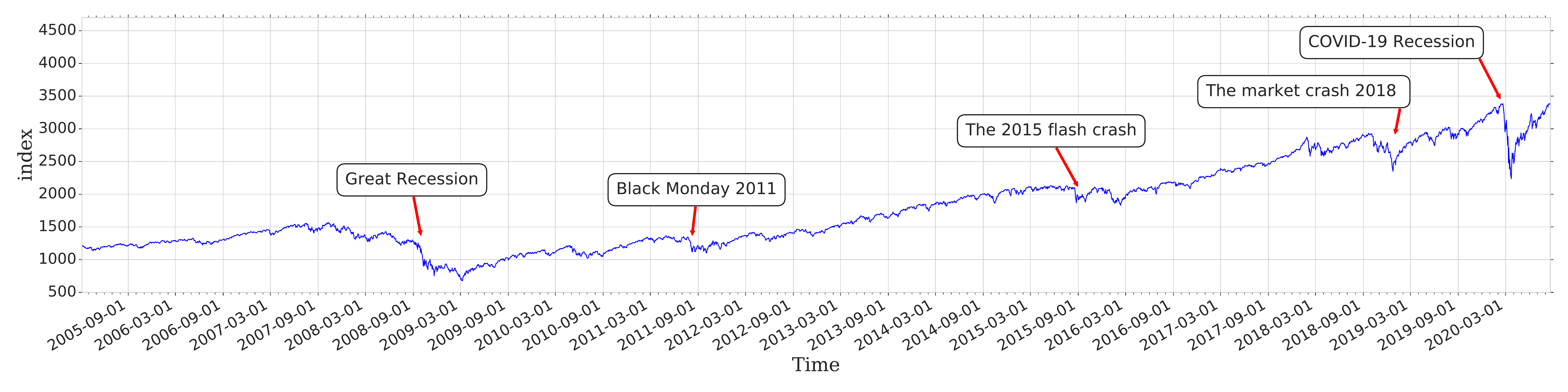}
	\caption{Time series of indexes obtained from $S\&P500$ with highlighting crises}
	\label{index}
\end{figure}

\begin{figure}[ht!]
	\centering
	\vspace{-0.5 cm }
	\hspace{-1 cm }
	\includegraphics[scale=.7]{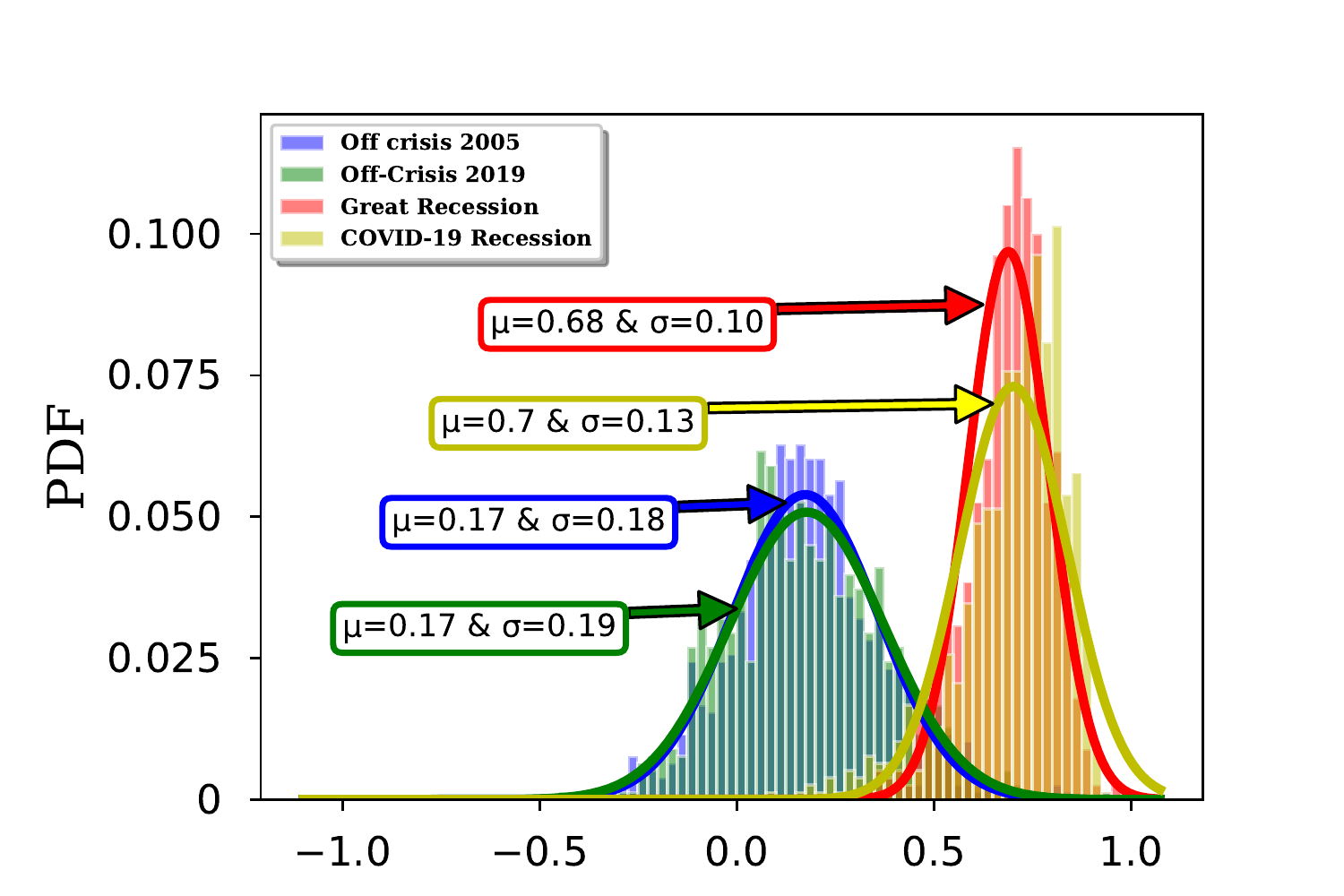}
	\caption{The probability distribution function (PDF) of correlation 
	matrix for four periods of time. Fitted curves to off-crisis 2005(blue), 
	off-crisis 2019(green) both have approximately zero mean value which is 
	representation of independent and random behavior of stocks in 
	off-crises periods. The positive mean values of curves in the 
	on-crises periods, Great Recession 2008(red), Covid-19 Recession 
	2019(yellow), demonstrate the strong correlation between stocks.}
	\label{probability}
\end{figure}

Financial markets are made by buying and selling numerous stocks and other 
types of financial instruments. Therefore, the statistical behavior of 
stocks that describes the micro-states of a market in different periods 
can shed light on the macro-states that emerge in the market. Crisis as an 
unforeseeable financial event in which the value of assets will drop 
significantly demonstrates that the fluctuation in the value of stocks 
percolates all over the market. A financial network is a complex network 
which its nodes represent the stocks, and links represent the correlation 
between stocks which can be used to predict the global properties of the 
financial market.  We used closed price of daily return of 40 companies of 
$S\&P 500$ index, from 2005 to 2020. We randomly choose 4 companies from 
10 distinct industries to investigating the effect of crisis overly. The 
regulated closing price of stocks is prepared by yahoofinance 
\cite{QuoteMedia}. We analyze four specific time widows. 
October 1,2008 - December 10,2008 
and January 29,2020 - April 8,2020 are close to financial crisis and we 
call them on-crisis 2008  and 2020. Also, July 28,2005 - October 
6,2005 and November 15,2018 - January 30,2019 are far from financial 
crisis and are called off-crisis 2005 and 2019 respectively. During the 
periods which we have selected there are two prominent financial crises the 
great recession in 2008\cite{namaki2, bahrami}, and the COVID-19 recession 
in 2020\cite{wheelock}. Fig.\ref{index} 
illustrate the behavior of the indexes over time. Different crises have 
marked in the figure and the two well-known crises 
that we want to address are the 2008 Global Financial crisis, Great 
Recession, and an ongoing global economic recession in direct result of 
the COVID-19 pandemic, COVID-19 Recession.
We use correlation matrix of stocks in which each element valued between 
$[-1,+1]$ and represents the correlation between two stocks. For 
correlation matrix we use the log-return of 
time-series $X_i(t)$ :
\begin{equation}\label{eq1}
	x_i(t) = \log X_i(t)-\log X_i(t-1),
\end{equation} 
where $x_i(t)$ is the corresponding log-return of the daily closing stock 
price of company $i$ at time $t$, $X_{i}$,. Now we can extract 
the correlation among stocks in time period $(t, t+\tau)$,  
in which $\tau$ is $50$ ( we have 77 time window that each window is for a 
period of 50 days ). For each time window, we calculate the correlation 
matrix 
$\mathbf{C}$, 
\begin{equation}\label{eq2}
	C_{ij} = \frac{1} {\sigma_i \sigma_j} \left< (x_i(t)-\mu_i) 
	(x_j(t)-\mu_j) \right>_{win(t, t+\tau)},
\end{equation}
where $C_{ij}$ is an element of the correlation matrix representing the 
relation between two stocks $i,j$ and $\langle\dotsb\rangle_{win(t, t+\tau)}$ 
is the time average over the 
time window from $t$ to $t+\tau$, and $\mu_{i}$ is the average of 
$x_i(t)$ over the time window defined as $\langle{x_i(t)} 
\rangle_{win(t,t+\tau)}$, and $\sigma_{i}$ is the standard 
deviation of $x_i(t)$ defined as $\sqrt{\langle (x_i(t)-\mu_i)^2 
\rangle_{win(t,t+\tau)}}$. Fig.~\ref{probability} displays the probability 
distribution function (PDF) of elements of the correlation matrix for four periods of on and off crisis which we mentioned. Simply, it can be 
seen that the PDF of elements of correlation matrix in off-crises and on-crises are 
different. The fitted Gaussian curves demonstrate that in both off-crises periods and on-crises periods markets behave almost similar as such the 
mean values, $\mu$, and the standard deviations, $\sigma$ are equal. The 
zero mean value of PDF in off-crises represents that the stocks have a 
random behavior and there is no significant correlation between them while 
positive non-zero mean values in on-crises periods is a confirmation of 
existence of a strong correlation between stocks which result in an 
emergence of a collective behavior of stocks in crisis.

\begin{figure}[ht]
	\centering 
	\hspace{-1 cm}
	a)\includegraphics[scale=.265]{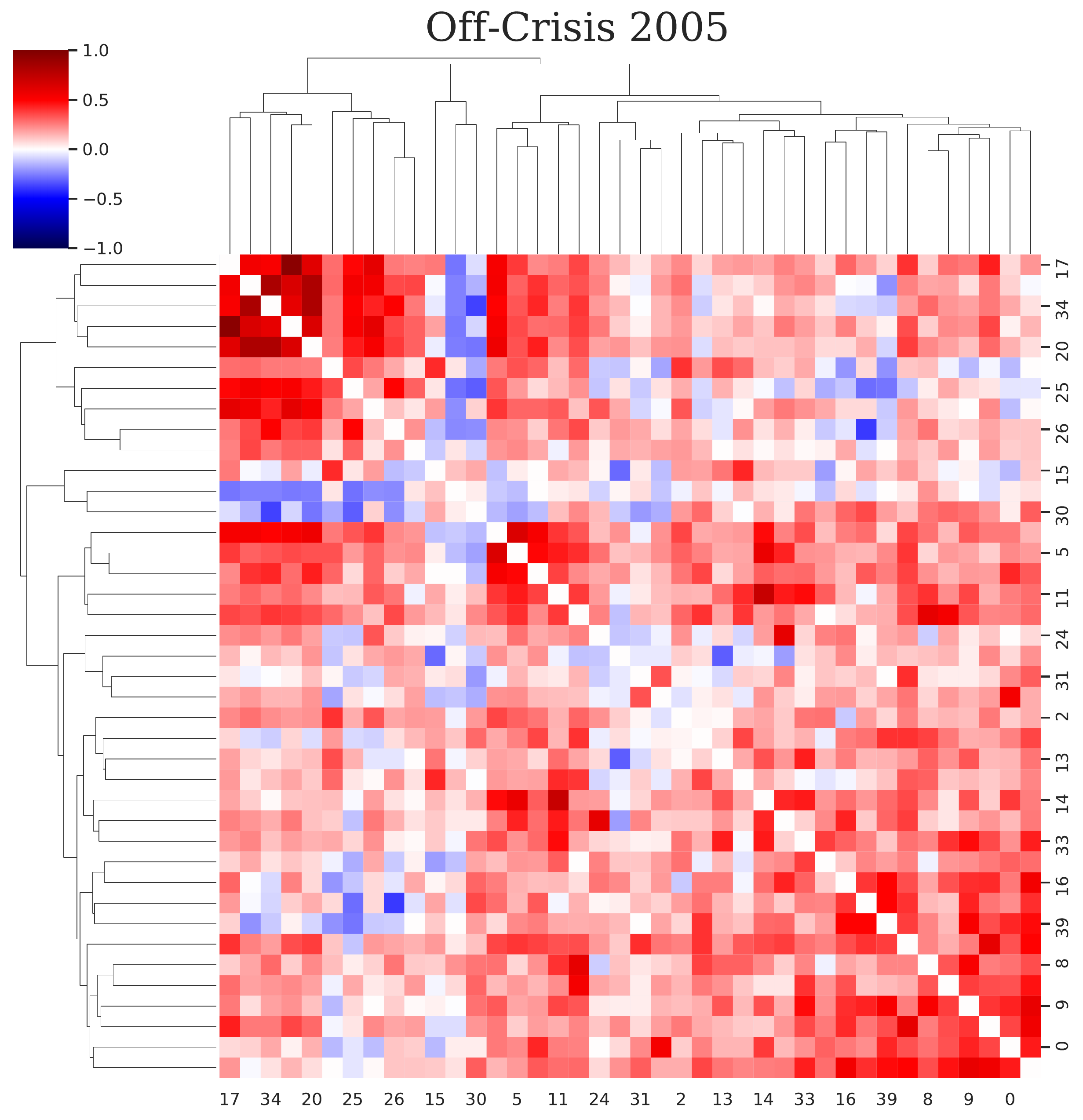}
	\hspace{1.5cm}
	b)\includegraphics[scale=.27]{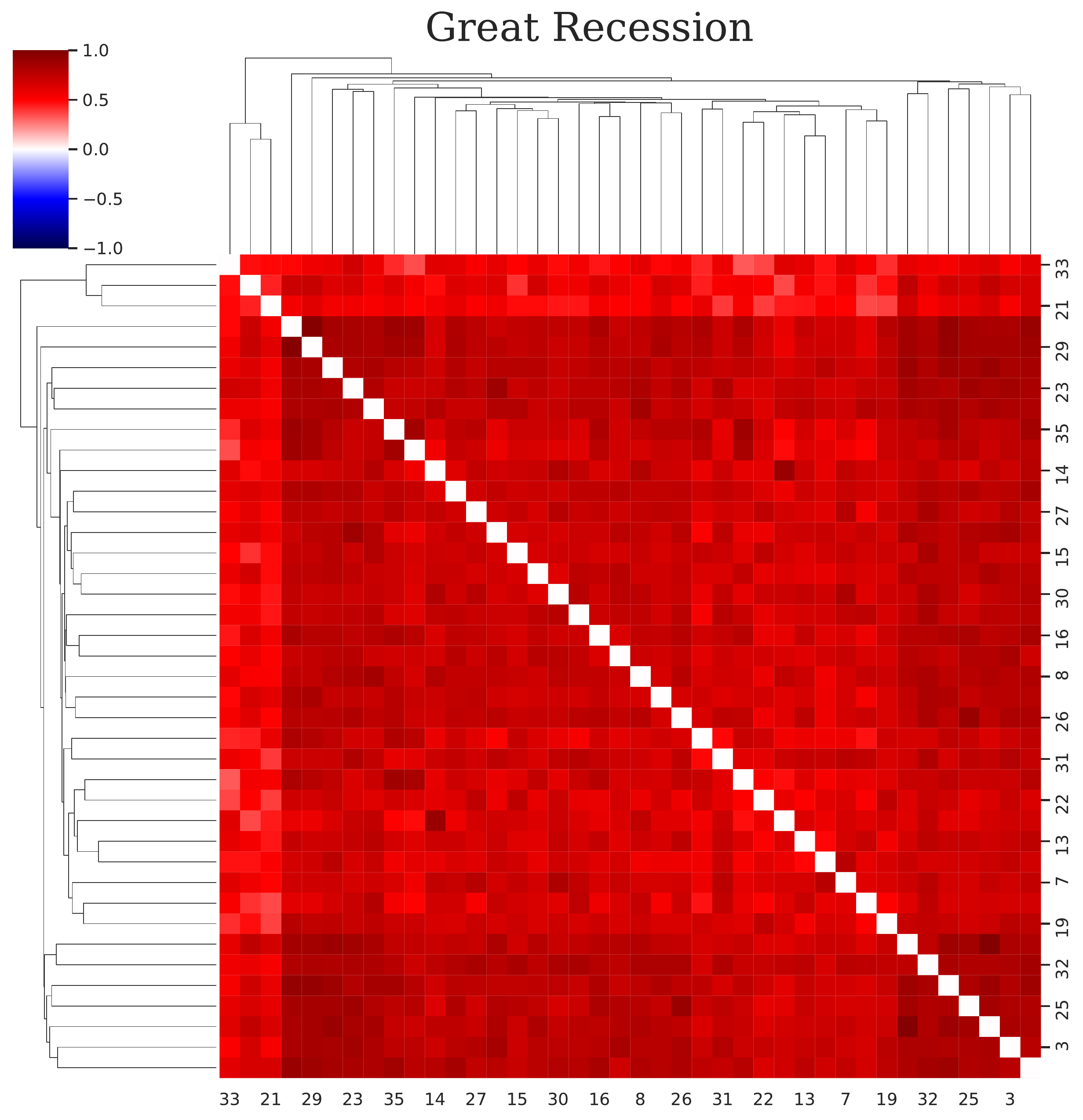}\\
	\vspace{0.5cm}
	\hspace{-1 cm}
	c)\includegraphics[scale=.265]{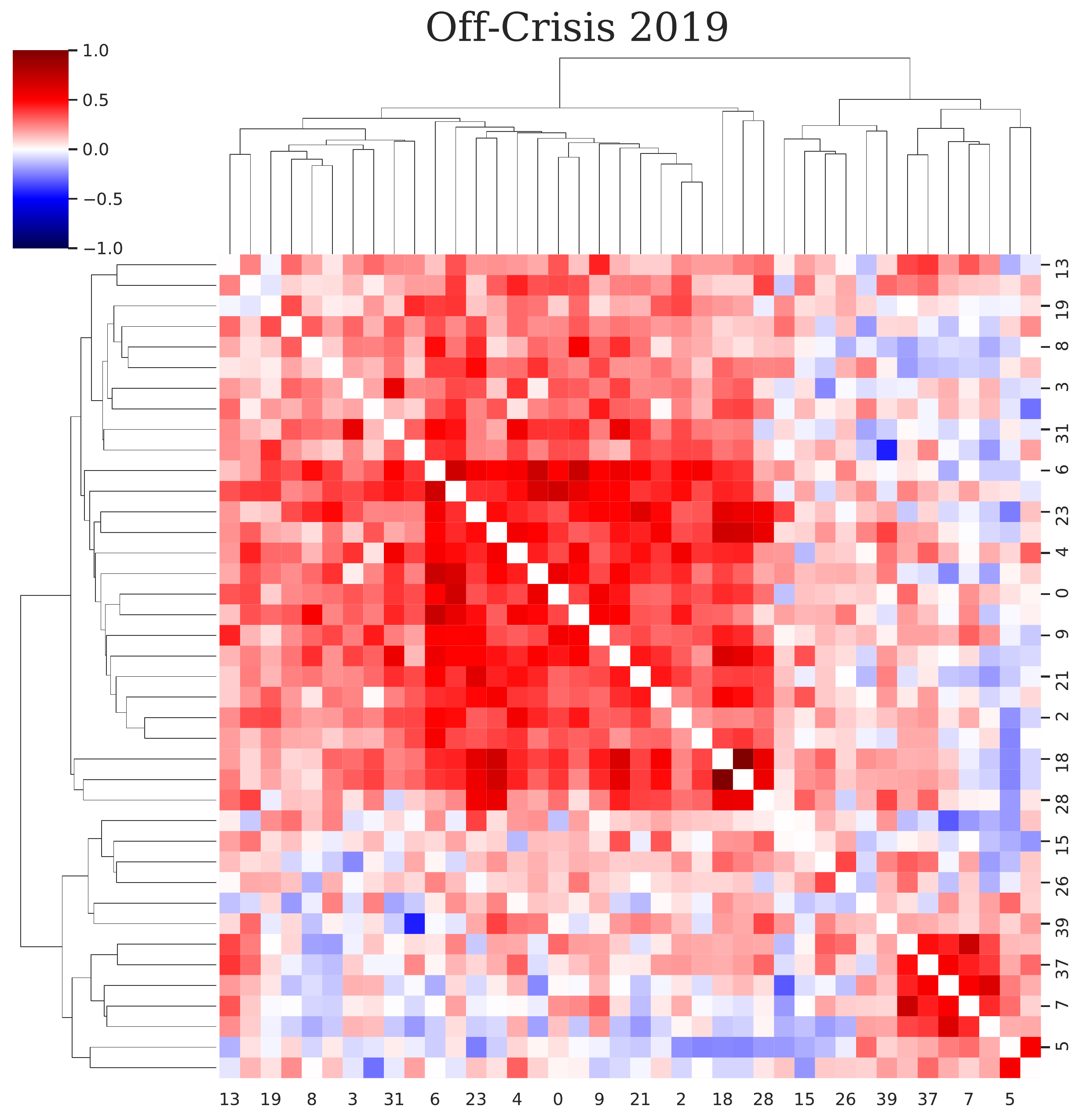}
	\hspace{1.5cm}
	d)\includegraphics[scale=.27]{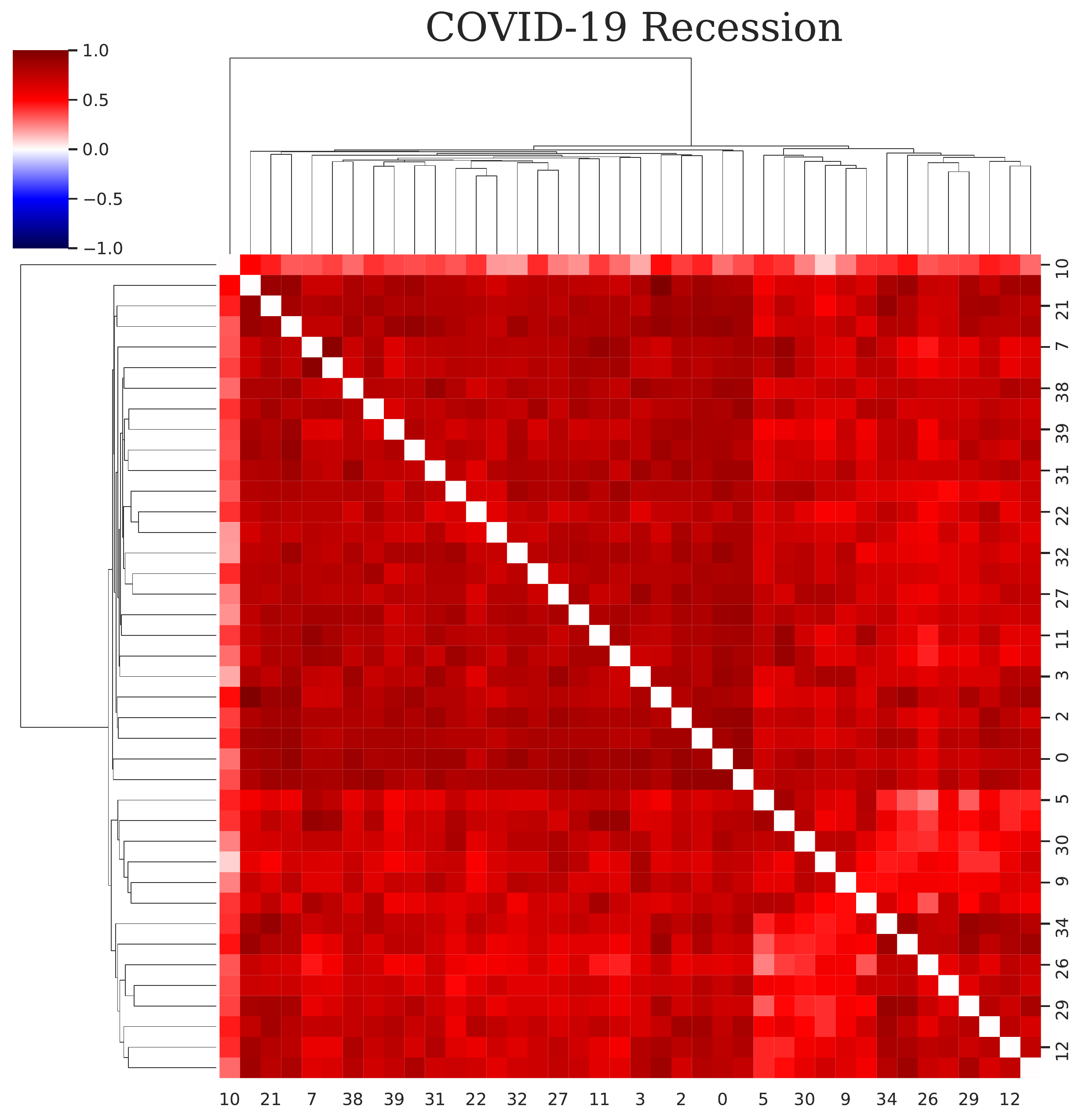}
	\caption{Heatmap of correlation matrix in a,c) Off-crisis and 
		b,d) On-crisis periods. We use dendrogram to cluster matrix elements 
		to illustrate the similarity of those. There is no 
		noticeable correlation between stocks in off-crisis (a,c) while in 
		on-crisis (b,d) correlation matrix gets block diagonal and we see the 
		strong correlation of the stocks.}
	\label{crisis12}
\end{figure}

In order to clarify how stocks are correlated in on and off crisis 
periods we plot the heat-map of correlation matrix. We sorted the 
correlation matrix elements in a way to group the similar values. 
Fig.~\ref{crisis12} upper row reveals the clustermap of correlation matrix 
in periods off-crisis 2005 and on-crisis 2008, Great Recession, and lower row is the 
clustermap of correlation matrix in periods off-crisis 2019 and on-crisis 
2019, COVID-19 Recession. In off-crisis cases Fig.\ref{crisis12}.(a,c) stocks are behaving independently far from the crisis so correlation 
values are random but in on-crises Fig.\ref{crisis12}.(b,d) the matrix 
gets block diagonal which reflects that stocks get correlated on-crises 
periods and represent a collective behavior of stocks. This means that 
fluctuation in value of stocks in a crisis are inter correlated and occur 
simultaneously/sequentially. 

Calculating the correlation matrix of stocks enable us to construct the 
financial network in which nodes represent stocks and each link states the 
correlation value between two nodes. In the quest of studying the effect 
of higher order interactions on the dynamics of the financial markets, we 
construct the network based on triadic relationships. Hence, we propose a Hamiltonian of triangles which links are elements of the correlation matrix and have continuous values. Inspired by the Hamiltonian proposed for balance theory which addresses a network of weighted triangles\cite{bagheri} here links are weighted in contrary to balance theory that each link can be either $+1$ or $-1$. Thus, the Hamiltonian is: 
\begin{equation}\label{eq3}
	H = - \sum_{i>j>k}{{{C}_{ij}}{{C}_{jk}}{{C}_{ki}}},
\end{equation}
in which ${{C}_{ij}}$ is an element of correlation matrix that indicates a 
link between two stocks $i$ and $j$. According to Eq.~(\ref{eq3}), the 
energy of the network is proportional to summation of multiplication of 
the triad's links. If the multiplication of links of a triangle is 
positive it is in balanced state and results in decreasing the energy of 
the system. The opposite is true for unbalanced triangles.
In the on-crisis period, the stocks are highly correlated and an ordered 
structure emerges in the network which indicates the formation of a big 
and strong community of correlated stocks. Imposing 
a disorder, interpreted as temperature, can measure the resistance of the 
ordered phase against changes. In low temperature, the system 
insists to keep its structural order while there is a 
critical temperature, $T_{c}$, that for temperatures above, a failure 
occurs in the structure of the network. So long as the correlation between 
stocks is strong, disorder can not overcome the ordered phase so, a higher 
temperature is needed to destroy the ordered structure. In the off-crisis period, 
markets are independent therefore the structure of the network does not 
show the existence of collective behavior. In this case, the critical 
temperature is approximately zero due to a lack of strong correlation between stocks. Investigating the dynamics of a financial 
network under various temperature is the main focus of following section.

\section*{Triadic Interactions under Balance theory in non-zero 
Temperature}
We proposed a Hamiltonian for the constructed network that considers 
triadic interactions. Since we know the Hamiltonian of the system, we can 
calculate the statistical quantities by Boltzmann-Gibbs 
distribution $p \sim e^{-\beta H(x)}$, where $H(x)$ is the Hamiltonian. Boltzmann-Gibbs distribution help us to derive all 
possible moments of the system to investigate their statistical behavior in 
equilibrium. The statistical mechanics in addition to 
mean-field approximation is our desired approach to understand the 
dynamics of the system statistically. In general, three quantities can be 
calculated, the mean value of the link, the two-star (two links with a common 
node), and the energy which is the mean value of the Hamiltonian. Using 
equilibrium statistical mechanic and mean-field method, we need to derive 
partition function. For a canonical ensemble that is classical and 
discrete, the canonical partition function is defined as \textit{$Z = 
\sum_{i} e^{-\beta E_{i}}$} 
where $\textit{i}$ is the index for microstates of the system, $\beta$ is 
the thermodynamic quantity proportional to the inverse of the temperature 
\textit{$1/T$}, and \textit{$E_{i}$} is the total energy of the system in 
the respective microstate. The system under study constructed based on the correlation matrix which its elements are weighted and reflect the dependency of stocks to each other; encourages us to use the weighted balance theory model that has been introduced in \cite{bagheri} comprehensively. Here by applying the following changes we can 
correspond the Hamiltonian \ref{eq3} with one being studied in \cite{bagheri}. The multiplication of the absolute value of correlation matrix elements represent the weight of triangles, $J_{ijk} = 
\mathopen|C_{ij}\mathclose|*\mathopen|C_{jk}\mathclose|*\mathopen|C_{ki}\mathclose|$, and the sign of each component represent a link of the network, $S_{ij} = \sgn(C_{ij})$. Via the applied changes the Hamiltonian Eq.\ref{eq3} can be rewritten in the following form: 
\begin{equation}\label{eq4}
H = -{\sum_{i>j>k}J_{ijk}S_{ij}S_{jk}S_{ki}},
\end{equation}	
where $J_{ijk}$ is the weight of a triangle has built on nodes $ i, j, k$ 
and $S_{ij}$ is the link between nodes $i, j$ that can be either $-1$ or 
$+1$. Using mean-field approach enable us to calculate the statistical quantities of the system.
\subsection{- Mean value of link}
According to mean-field method the Hamiltonian Eq.\ref{eq4} can be divided into two parts, $H = H_{ij} + H'$,
 where ${H}_{ij}$ includes all terms in the Hamiltonian that contain $S_{ij}$, and $H'$ is the rest of the Hamiltonian that relates to other links, $\{S'\}$, \cite{bagheri}.
\begin{equation}\label{eq5}
\begin{aligned}[b]
&H_{ij}=- S_{ij}{\sum_{k\neq{i,j}}J_{ijk}S_{jk}S_{ki}},\\
&p\equiv\langle S_{ij}\rangle =\sum_{S_{ij}=\{\pm{1}\}} S_{ij} \ P(S_{ij})= (1)\times P(S_{ij}=1)+ (-1)\times P(S_{ij}=-1)   =\frac{1}{Z}\sum_{\{S'\}}e^{-\beta H'}\sum_{S_{ij}={\left\{\pm{1}\right\}}}S_{ij}e^{-\beta
	H_{ij}}	\\
&p = \tanh(\beta\langle{\sum_{k\neq{i,j}}J_{ijk} \ S_{jk} \ S_{ki}}\rangle),
\end{aligned}
\end{equation} 
where $Z = \sum \exp{(-\beta H) }$ is the partition function and the term in the parentheses is called the mean value of weighted two-stars, $Q\equiv\langle 
\sum_{k\neq{i,j}}J_{ijk}S_{jk}S_{ki}\rangle$, and is interpreted as the mean-field each link feels.
\subsection{- Mean value of two-star}
Following the same process mentioned above the Hamiltonian can be divided in the form of, $ H = H_{jk,ki} + H''$, where $ H_{jk,ki}$ consists of all term including $S_{jk}$ and $S_{ki}$, and $H''$ is the rest, \cite{bagheri}. 
\begin{equation}\label{eq6}
\begin{aligned}[b]
& H_{jk,ki}=-S_{jk}(\sum_{l\neq{i,j,k}}J_{jlk}S_{jl}S_{kl})-S_{ki}(
\sum_{l\neq{i,j,k}}J_{kil}S_{kl}S_{il})-J_{ijk}S_{ij}S_{jk}S_{ki},\\
& q\equiv \langle S_{jk}S_{ki} \rangle = \sum_{S_{jk},S_{ki} = \{\pm{1}\}} S_{jk} S_{ki} \ P(S_{jk}S_{ki}) = \frac{e^{2\beta Q}-2\,e^{-2\beta J_{ijk} \tanh(\beta Q)} 
	+e^{-2\beta Q}}{e^{2\beta Q}+2\,e^{-2\beta J_{ijk} 
		\tanh(\beta Q)} 
	+e^{-2\beta Q}} =  q(Q,J_{ijk},\beta),
\end{aligned}
\end{equation} 
the fraction represents that $q$ is the function of weighted mean value of two-star, $Q$, temperature , and the weights, $J_{ijk}$ that has been assumed are coming from Gaussian probability distribution. The detailed process of preceding analytical solutions have been completely done in \cite{bagheri}. Considering continuous values of weights,  we change the summation in the definition of $Q$ to integral, $\sum_{k\neq{i,j}} \rightarrow \int_{-\infty}^{\infty} dJ \ P(J)$, and we integrate over all weights that are coming from Gaussian 
probability distribution to calculate the mean field each link feels due to weighted two-stars over it.
\begin{equation}\label{eq7}
Q \equiv \langle \sum_{k\neq{i,j}}J_{ijk}S_{jk}S_{ki}\rangle =   
(N-2)\int_{-\infty}^{\infty}  J' \  P(J') \ 
q(Q,J',\beta) \ 
dJ' = f (Q, \mu, \sigma, N, \beta).
\end{equation} 
The result is a function of the mean value and the variance of the Gaussian probability distribution. The coefficient, $N-2$, is the normalization parameter; It is the number of two-stars on a specific link. While here we represent an analytical solution by the assumption of Gaussian probability distribution of weights, in working with real data the probability distribution of weights is inferred from data. Solution of the self-consistence equation \ref{eq7} confirms that there is a critical temperature, $T_{c}$, which determines the state of the system. For the temperatures above, $T > T_{c}$, the network is in a random state of positive and negative links whereas for the temperature below the critical, $T < T_{c}$, the network is in a balanced state whether paradise (all links are positive) or bipolar (division of network into two groups that are friendly relationship inside and enmity between groups). Therefor the system 
experiences a phase transition in the result of a sudden jump between the 
ordered structure and the random state. In other words, the resistance of the ordered structure against changes can be measured by critical temperature. Since stocks behave similarly in the crisis periods, an ordered structure forms that is resistance against disorder and higher temperature is needed for the network to make a transition. Therefore the stronger the crisis, the higher the critical temperature. Investigating the behavior of statistical quantities of the constructed network based on real financial data in different periods of time  is going to be discussed and illustrated in the next section. 

\section*{Critical Temperature on and off crisis}

Introducing a Hamiltonian that describes the dynamics of a system and 
considering a parameter, temperature $(T)$, that can be interpreted as 
tension and randomness allows a system to fluctuate and provides us the 
chance of studying a system by Boltzmann-Gibbs statistics. Theoretical 
solution and derivation of the self-consistence equation \ref{eq7} tells 
us there is a 
critical temperature, $T_{c}$, that network is in unbalanced state for 
temperature higher than $T_{c}$ and in balanced state for temperature 
lower than $T_{c}$ and the system experiences a first order phase 
transition. To confirm this result by simulation we consider the 
constructed network as the initial state of the system then in order to  
study the fully connected network of stocks we run a 
Metropolis Monte Carlo simulation on the initial network to study its 
evolution under different temperatures. In the simulation process, at each 
Monte Carlo step for any temperature, initially, we set all link signs 
randomly (it means $T = \infty$). In the following, we choose a link 
by chance and switch the sign in order to increase the number of balanced 
triads that is equivalent to energy reduction. If switching sign of the 
link lessens the energy of the network, we will accept this change and the 
sign of link is updated, otherwise, we will accept this change by a 
Boltzmann probability proportional to ${{exp}(-\beta \Delta E)}$ where 
$\Delta E$ is the energy difference, $(E_{f}-E_{i})$, after and before 
flipping the sign of randomly selected link. Independent of initial condition, for $T > T_{c}$ the structure is random while considering initial condition the state of the system will vary for $T < T_{c}$. In the case of random initial condition the balanced state of the network below critical temperature is bipolar therefore the mean value of weighted two-stars equals zero. For positive initial condition the network undergoes a dynamic transition to heaven for $T < T_{c}$ and the mean value of weighted two-stars equals one. The behavior of the order parameter, $Q$, versus temperature is a suitable quantity to illustrates the differences of on-crisis and off-crisis periods in a financial network.

\begin{figure}[ht!]
	\hspace{-0.8 cm}
	\includegraphics[scale=.45]{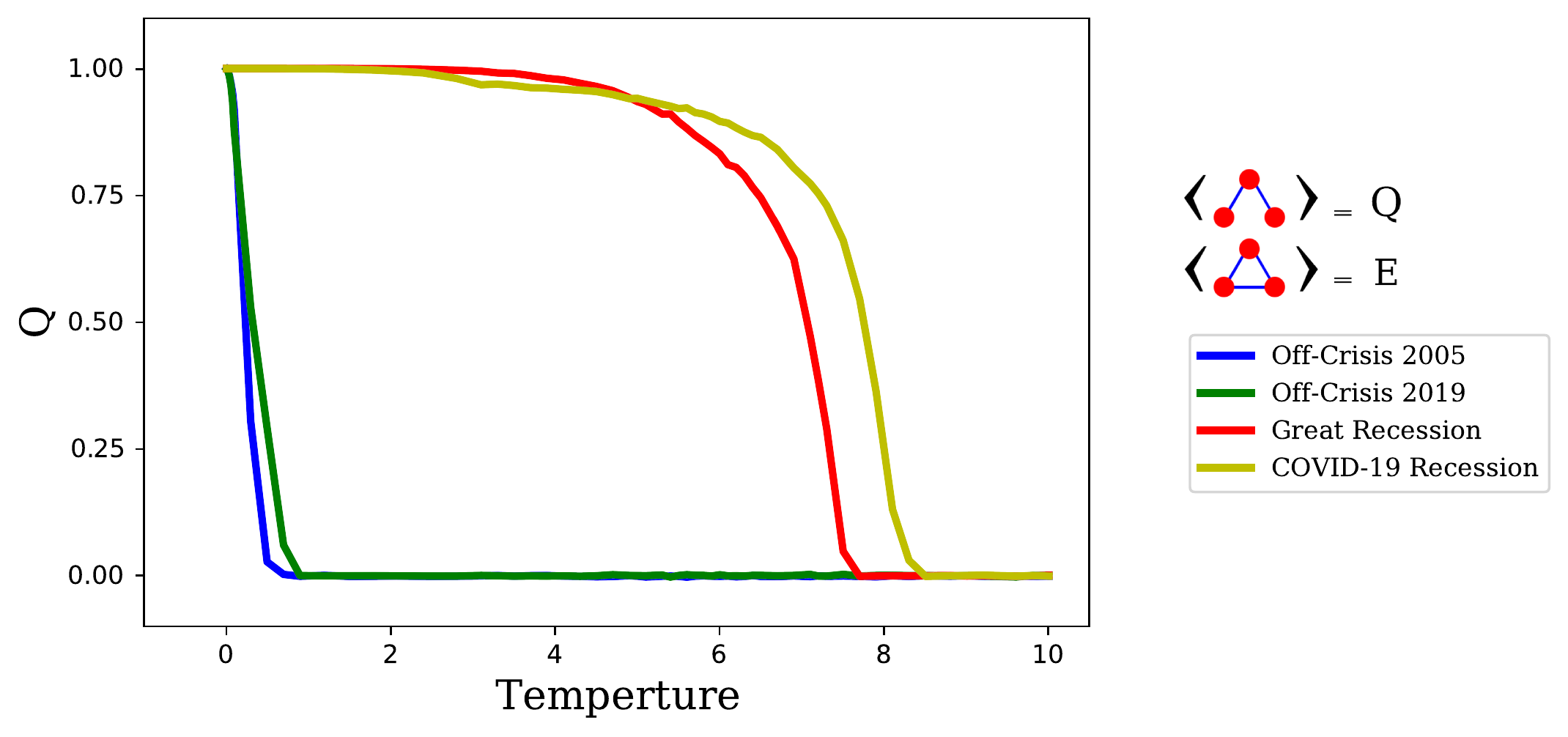}
	\includegraphics[scale=.45]{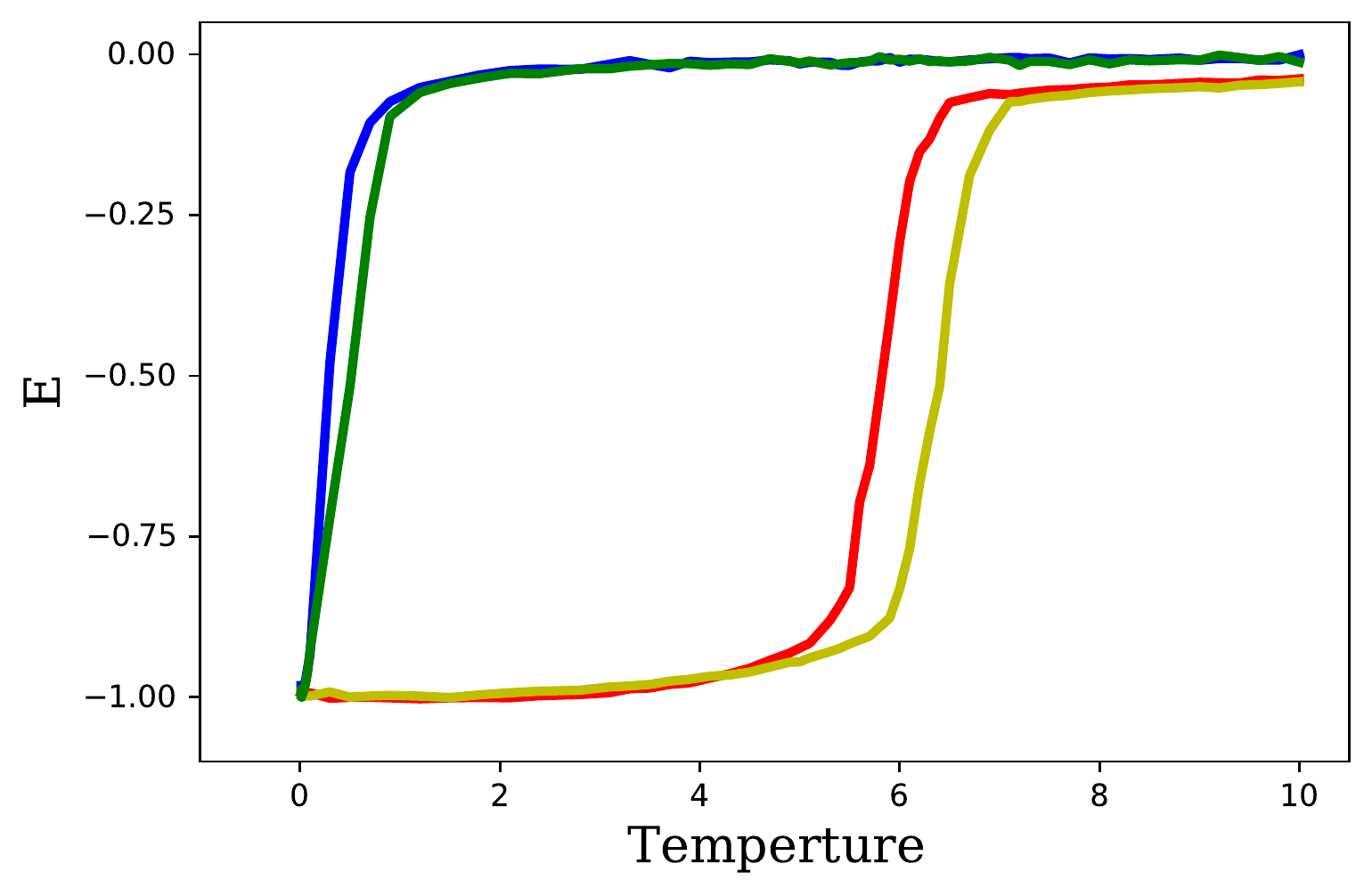}
	\caption{a) The mean value of weighted two-stars, $Q$, versus temperature in 
		two different on-crisis periods. The $Q$ equals $1$ in low temperature 
		while increasing temperature,$T$, will result in destroying the ordered 
		structure of the network and zero value of the $Q$ in the random 
		configuration of the network. In off-crisis periods the value of 
		$Q$ is different from zero around approximately zero temperatures and it will 
		quickly reduces to zero due to the weak correlation between stocks, 
		b) The energy of the network versus temperature. In 
		the on-crisis the structural order of the network resists against 
		changes hence it is needed to increase the temperature 
		to a critical value that can overcome the ordered phase. In off-crisis 
		periods the structure of the network is not in an ordered phase so 
		a small amount of disorder can push the network in a random 
		configuration and zero value of energy.}
	\label{order}
\end{figure}


\begin{figure}[ht!]
	\centering
	\vspace{0.5 cm}
	a) \includegraphics[scale=.53]{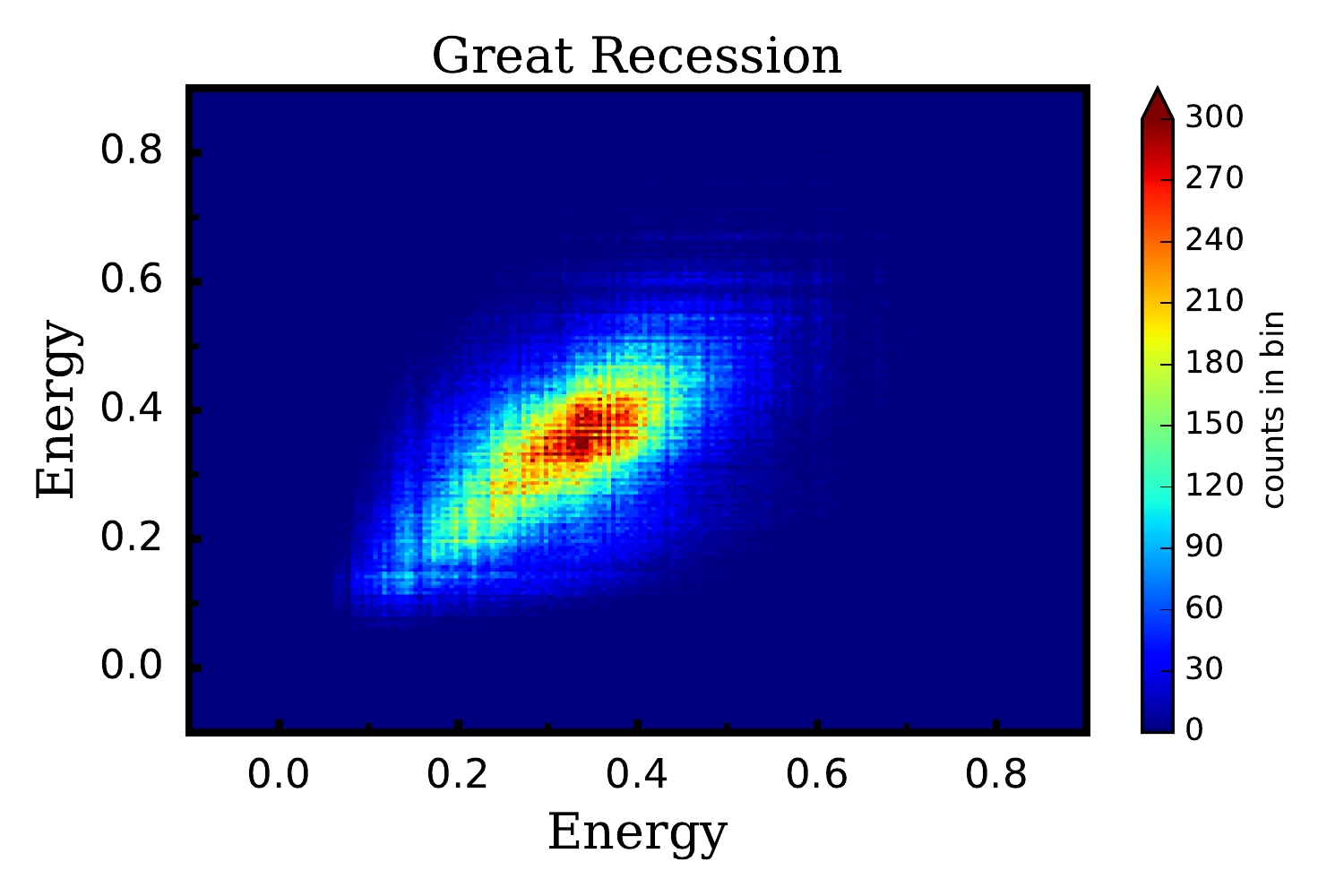}
	b) \includegraphics[scale=.53]{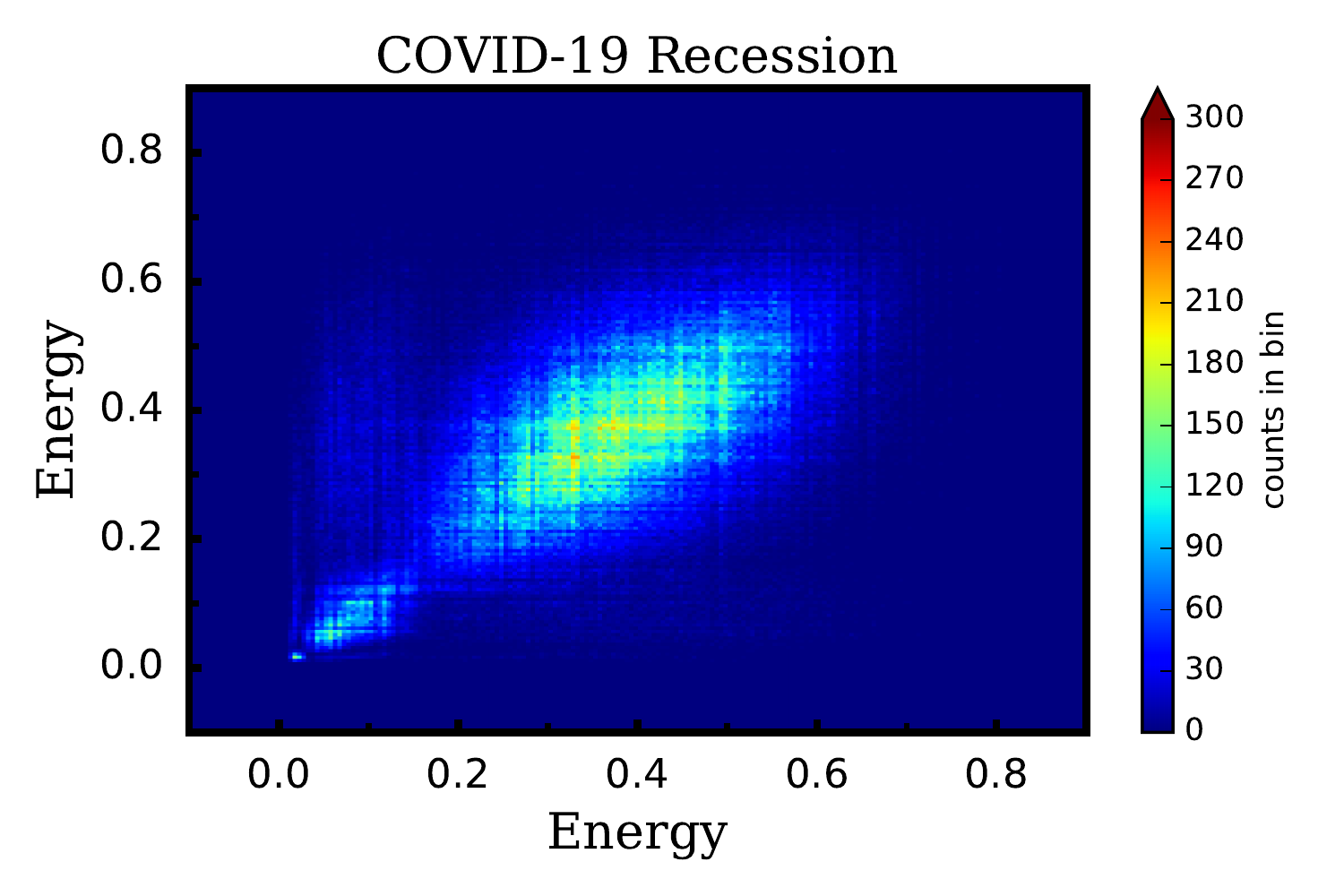}\\
	c) \includegraphics[scale=.53]{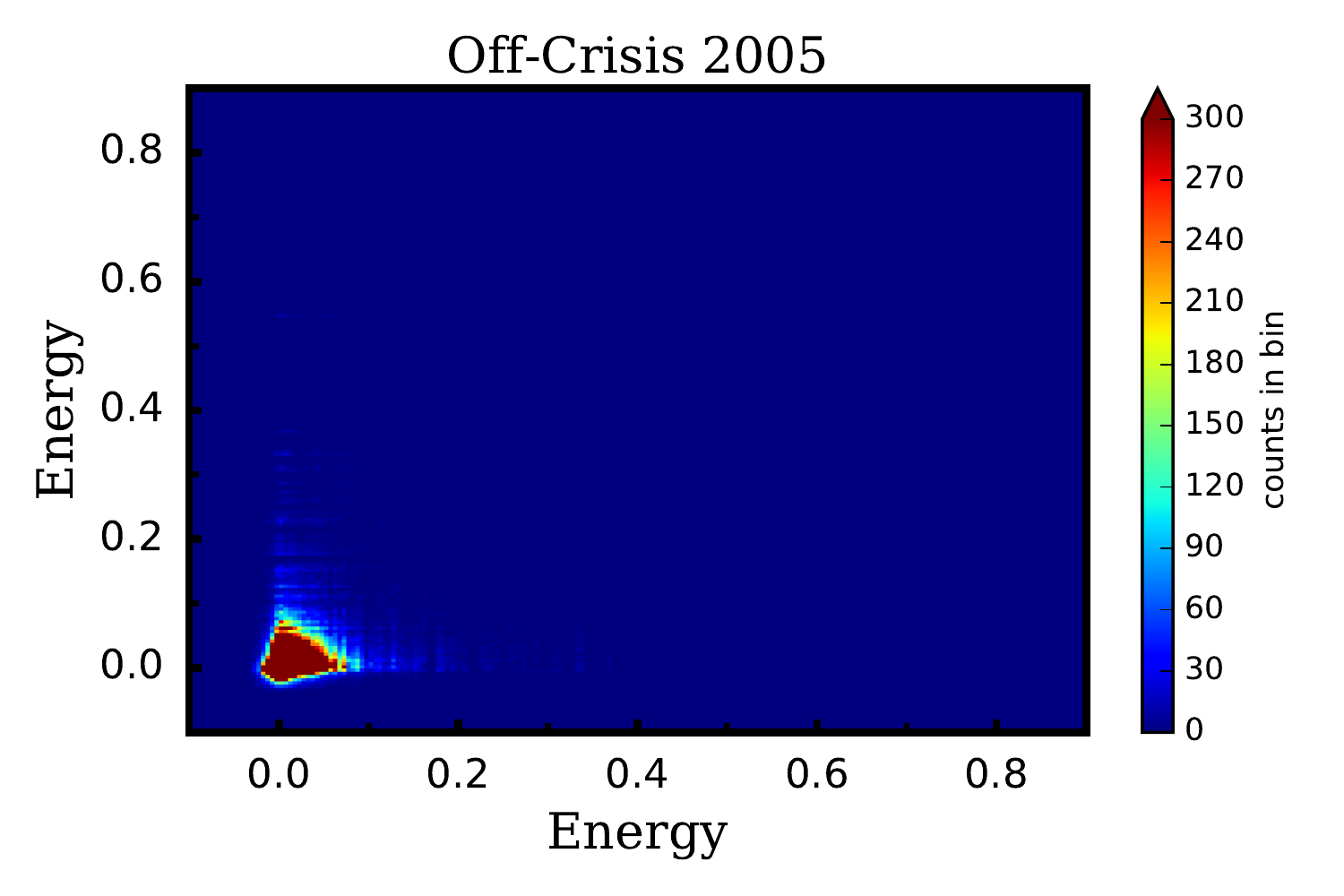}
	d) \includegraphics[scale=.53]{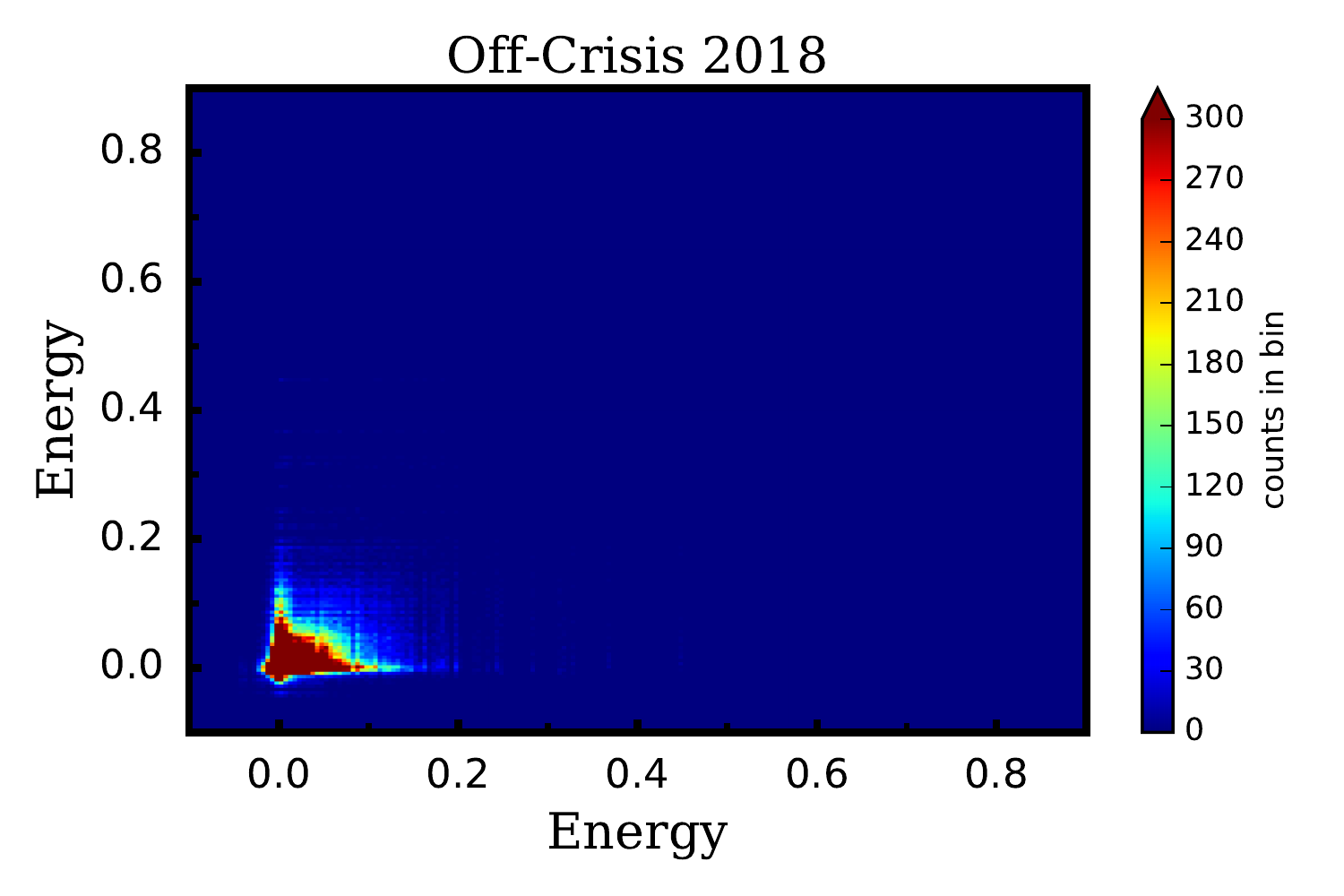}
	\caption{Upper row show the energy-energy landscape for a) the Great 
		recession and b) the COVID-19 recession. Lower row is the energy- 
		energy landscape for two off-crises periods c) July 28, 2005- 
		October 6, 2005 and d) November 15, 2018 –January 30, 2019. Each 
		point is a representative for the number of triangles with a 
		common link which have the specified vales of energies. In the 
		on-crises periods triangles with various energies are connected 
		while in off-crises periods points are localized to small 
		energies.}
	\label{crisis123}
\end{figure}

Fig.\ref{order}.(a) demonstrate the order parameter, $Q$ ( mean 
value of weighted two-star ), in on-crisis and off-crisis periods; the presence 
of strong correlation between stocks in the on-crisis indicates 
the existence of a collective behavior among stocks and formation of an 
ordered structure in the network. In the ordered phase $Q=1$ which 
states that all triangles in the network are in the balanced state (heaven). 
Applying disorder, increasing temperature, can not change the ordered 
structure until the point where the ordered phase of the network disrupts 
in a critical temperature, $T_{c}$, and its consequence is $Q=0$. In off-crisis 
periods, due to the lack of strong 
correlation between stocks, the critical temperature is approximately zero increasing 
temperature results in random structure of the 
network and zero value of $Q$. The collective behavior of the network in on-crisis preventing the system from going to random phase so higher value of critical temperature is reported in respect to off-crisis. Fig.\ref{order}.(b) illustrates the energy of the network versus temperature in 
two periods of on-crisis and off-crisis. In the ordered phase which all 
triangles of the network are in the balanced state the total normalized 
energy of the system equals $-1$ according to Eq.\ref{eq3}. Although the 
ordered phase preserve its structure even by applying disorder, there is a 
critical point, $T_{c}$, in which above the structure destroys abruptly. 
Increasing temperature leads the structure toward a random configuration 
of balanced and unbalanced triangles hence the energy of the network 
equals zero and the network is random. The difference in critical 
temperature of two on-crises periods states that the COVID-19 Recession is 
a stronger crisis in respect to Great Recession in 2008 since it needs 
more disorder to overcome the strong correlation of stocks. Notice that in Fig.\ref{order} a,b, temperature where the network make transition from ordered structure to random structure is reported differently.  Recalling that the elements of the correlation matrix are real values between $-1$, $1$ and
$E$ is multiplication of three elements clarify that energy has smaller amount relative to $Q$ which is multiplication of two elements. Hence the energy versus temperature Fig.\ref{order}.b, displays smaller critical temperature while our judgment is based in the order parameter, $Q$.

In the off-crisis periods their critical temperatures is near zero which means stocks 
are weakly correlated and behave independently so the energy of the network equals zero. Fig.~\ref{crisis123} shows 
the energy landscape of triangles that have 
a joint link. Each triangle, $u$, has specific energy, $E_{u}$ and if two 
triangles $u$ and $v$ share a common link, we would consider the 
$(E_{u},E_{v})$ as a point in the energy-energy plot. To clarify the 
method, we use 2D-histogram to count the number of triangles exist in the 
different area of energy-energy landscape. It is easy to see that in the 
on-crisis 2008 and on-crisis 2020, triangles with common link have more 
energy and wide pattern of connections. However, the pattern of 
connections of triangles is localized in the off-crisis periods. 
Note that, because of the localization of energy in the off-crisis, there 
are bins in which the abundance of triangles that have joint link is much 
more (80000 connected triangles) than in the on-crisis 
(300 connected triangles). So, for better illustration, we use an extended 
2D-histogram. Actually, bins with an abundance of more than 300 are 
colored red. \\

\begin{figure}[!ht]
	\centering
	\hspace{-0.3cm} 
	\includegraphics[scale=.175]{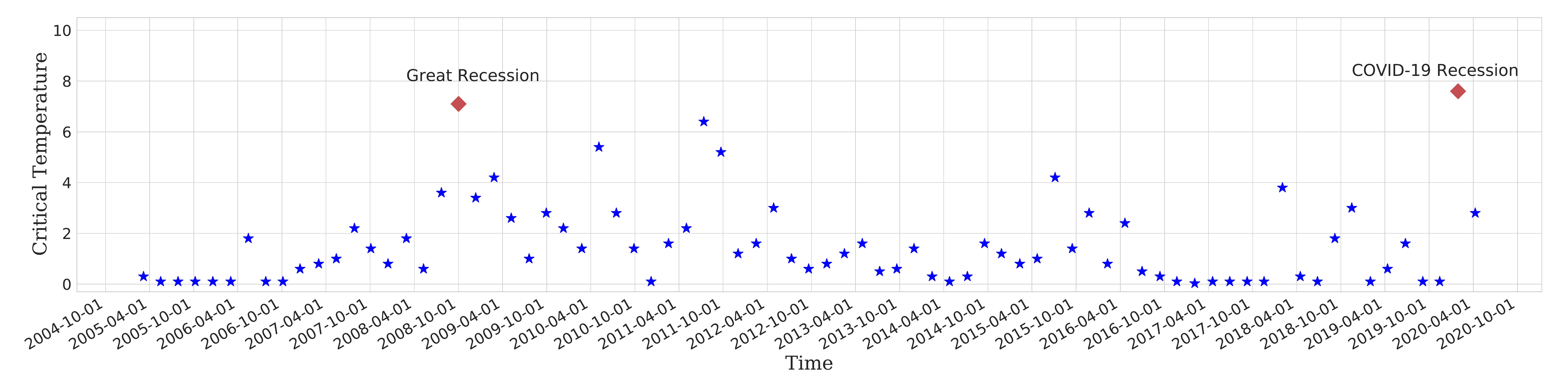}
	\caption{Critical temperature for different crises. The stronger 
	correlation of stocks results in a more dangerous crisis and a higher 
	disorder is needed to get out of it.}
	\label{critical}
\end{figure}

Fig.~\ref{critical} shows critical temperature for several 
time windows in 
which presence of correlation leads to increase the critical temperature. 
Actually, the collective behavior of stocks results in an ordered 
structure that is resistant to changes hence higher temperature is needed 
to overcome this behavioral similarity of stocks. Note that, in the 
figure, we highlight the time window of outstanding crises in our selected 
time period (2005-2020) by red diamond. One can see that the critical 
temperature in the crisis time windows is more than windows that are far 
from crisis.

\section*{Conclusion and Discussion}
The connection between links ( stocks correlation ) and the complex behavior of financial networks in crisis has been devoted our attention to the essential role of
higher order interactions. The significance of higher order interactions lies in order parameters that enables us to study such a this complex system statistically. The phase transition of the system in off-crises periods and their near zero critical temperature reports that applying temperature result in the random structure of the network and stocks behave independently. In addition, the strong connection between links in crisis indicates that an ordered structure forms in the financial network that we take advantages of a series of order parameters to describe this phase in the framework of balance theory. The order parameters state that an ordered structure has been construct in the network during on-crises periods that resists against applying disorder, temperature. The functionality of the order parameter, Q, 
versus temperature, T, displays that in on-crisis periods resistivity of the network 
will increase and the system has higher critical temperature, $T_{c}$ ( 
$T_{c}(Great Reseccion)\approx T_{c}(COVID-19 Recession)$) in respect to 
off-crisis. Even comparing the critical temperate of two on-crisis periods tells us that ordered structure formed during COVID-19 Recession is more resistance respect to Great Recession since it has higher critical temperature. More ordered structure, higher $T_{c}$; this is the main point that refers to temperature as a quantity that can measure the strength of a crisis. This evidence brings us to conclude that higher order interactions have key contribution in the realization of on-crisis periods via network structure.

\bibliography{Myreference}

\section*{Author contributions statement}
M.Z., M.B., and G.R.J conceived and designed the study. M.Z, M.B, and G.R.J conducted the experiments. M.Z. and M.B. analyzed the data, performed the statistical analysis,  M.Z created the figures, and M.B wrote the first draft of the manuscript. M.Z., M.B., A.T and G.R.J analyzed the results. All authors read and approved the final manuscript.

\section*{Competing interests}
The authors declare no competing interests.

\end{document}